\documentclass[submission]{eptcs}
\providecommand{\event}{GandALF 2017} 
\usepackage{breakurl}             
\usepackage{underscore}           

\usepackage{xparse}
\usepackage{amssymb}
\usepackage{MnSymbol}
\usepackage{stmaryrd}
\usepackage{graphics}
\usepackage[ruled, vlined]{algorithm2e}

\usepackage{amsthm}

\usepackage{microtype}

\NewDocumentCommand\A{g}{\IfNoValueTF{#1}{\textbf{A}}{\textbf{A}~#1}}
\NewDocumentCommand\E{g}{\IfNoValueTF{#1}{\textbf{E}}{\textbf{E}~#1}}

\NewDocumentCommand\G{g}{\IfNoValueTF{#1}{\textbf{G}}{\textbf{G}~#1}}
\NewDocumentCommand\F{g}{\IfNoValueTF{#1}{\textbf{F}}{\textbf{F}~#1}}
\NewDocumentCommand\X{g}{\IfNoValueTF{#1}{\textbf{X}}{\textbf{X}~#1}}
\NewDocumentCommand\U{gg}{\IfNoValueTF{#1}{\textbf{U}}{#1~\textbf{U}~#2}}
\NewDocumentCommand\W{gg}{\IfNoValueTF{#1}{\textbf{W}}{#1~\textbf{W}~#2}}

\NewDocumentCommand\EG{g}{\IfNoValueTF{#1}{\textbf{EG}}{\textbf{EG}~#1}}
\NewDocumentCommand\EF{g}{\IfNoValueTF{#1}{\textbf{EF}}{\textbf{EF}~#1}}
\NewDocumentCommand\EX{g}{\IfNoValueTF{#1}{\textbf{EX}}{\textbf{EX}~#1}}
\NewDocumentCommand\EU{gg}{\IfNoValueTF{#1}{\textbf{EU}}{\textbf{E[}#1~\textbf{U}~#2\textbf{]}}}
\NewDocumentCommand\EW{gg}{\IfNoValueTF{#1}{\textbf{EW}}{\textbf{E[}#1~\textbf{W}~#2\textbf{]}}}

\NewDocumentCommand\AG{g}{\IfNoValueTF{#1}{\textbf{AG}}{\textbf{AG}~#1}}
\NewDocumentCommand\AF{g}{\IfNoValueTF{#1}{\textbf{AF}}{\textbf{AF}~#1}}
\NewDocumentCommand\AX{g}{\IfNoValueTF{#1}{\textbf{AX}}{\textbf{AX}~#1}}
\NewDocumentCommand\AU{gg}{\IfNoValueTF{#1}{\textbf{AU}}{\textbf{A[}#1~\textbf{U}~#2\textbf{]}}}
\NewDocumentCommand\AW{gg}{\IfNoValueTF{#1}{\textbf{AW}}{\textbf{A[}#1~\textbf{W}~#2\textbf{]}}}

\NewDocumentCommand\CE{gg}{\IfNoValueTF{#1}{\ensuremath\llangle\rrangle}{\IfNoValueTF{#2}{\ensuremath\llangle#1\rrangle}{\ensuremath\llangle#1\rrangle~#2}}} 
\NewDocumentCommand\CA{gg}{\IfNoValueTF{#1}{\ensuremath\llbracket\rrbracket}{\IfNoValueTF{#2}{\ensuremath\llbracket#1\rrbracket}{\ensuremath\llbracket#1\rrbracket~#2}}} 

\NewDocumentCommand\CEG{gg}{\IfNoValueTF{#1}{\ensuremath\llangle\Gamma\rrangle\textbf{G}}{\IfNoValueTF{#2}{\ensuremath\llangle#1\rrangle\textbf{G}}{\ensuremath\llangle#1\rrangle\textbf{G}~#2}}}
\NewDocumentCommand\CEF{gg}{\IfNoValueTF{#1}{\ensuremath\llangle\Gamma\rrangle\textbf{F}}{\IfNoValueTF{#2}{\ensuremath\llangle#1\rrangle\textbf{F}}{\ensuremath\llangle#1\rrangle\textbf{F}~#2}}}
\NewDocumentCommand\CEX{gg}{\IfNoValueTF{#1}{\ensuremath\llangle\Gamma\rrangle\textbf{X}}{\IfNoValueTF{#2}{\ensuremath\llangle#1\rrangle\textbf{X}}{\ensuremath\llangle#1\rrangle\textbf{X}~#2}}}
\NewDocumentCommand\CEU{ggg}{\IfNoValueTF{#1}{\ensuremath\llangle\Gamma\rrangle\textbf{U}}{\IfNoValueTF{#2}{\ensuremath\llangle#1\rrangle\textbf{U}}{\ensuremath\llangle#1\rrangle\textbf{[}#2~\textbf{U}~#3\textbf{]}}}}
\NewDocumentCommand\CEW{ggg}{\IfNoValueTF{#1}{\ensuremath\llangle\Gamma\rrangle\textbf{W}}{\IfNoValueTF{#2}{\ensuremath\llangle#1\rrangle\textbf{W}}{\ensuremath\llangle#1\rrangle\textbf{[}#2~\textbf{W}~#3\textbf{]}}}}

\NewDocumentCommand\CAG{gg}{\IfNoValueTF{#1}{\ensuremath\llbracket\Gamma\rrbracket\textbf{G}}{\IfNoValueTF{#2}{\ensuremath\llbracket#1\rrbracket\textbf{G}}{\ensuremath\llbracket#1\rrbracket\textbf{G}~#2}}}
\NewDocumentCommand\CAF{gg}{\IfNoValueTF{#1}{\ensuremath\llbracket\Gamma\rrbracket\textbf{F}}{\IfNoValueTF{#2}{\ensuremath\llbracket#1\rrbracket\textbf{F}}{\ensuremath\llbracket#1\rrbracket\textbf{F}~#2}}}
\NewDocumentCommand\CAX{gg}{\IfNoValueTF{#1}{\ensuremath\llbracket\Gamma\rrbracket\textbf{X}}{\IfNoValueTF{#2}{\ensuremath\llbracket#1\rrbracket\textbf{X}}{\ensuremath\llbracket#1\rrbracket\textbf{X}~#2}}}
\NewDocumentCommand\CAU{ggg}{\IfNoValueTF{#1}{\ensuremath\llbracket\Gamma\rrbracket\textbf{U}}{\IfNoValueTF{#2}{\ensuremath\llbracket#1\rrbracket\textbf{U}}{\ensuremath\llbracket#1\rrbracket\textbf{[}#2~\textbf{U}~#3\textbf{]}}}}
\NewDocumentCommand\CAW{ggg}{\IfNoValueTF{#1}{\ensuremath\llbracket\Gamma\rrbracket\textbf{W}}{\IfNoValueTF{#2}{\ensuremath\llbracket#1\rrbracket\textbf{W}}{\ensuremath
\llbracket#1\rrbracket\textbf{[}#2~\textbf{W}~#3\textbf{]}}}}

\title{A Backward-traversal-based Approach\\for Symbolic Model Checking of Uniform Strategies\\for Constrained Reachability}
\author{
	Simon Busard\thanks{This research is financed by the Walloon Region as part of the Logistics in Wallonia competitiveness pole.}
	\institute{
		ICTEAM Institute,\\
		Universit\'e catholique de Louvain,\\
		Louvain-la-Neuve, Belgium
	}
	\email{simon.busard@uclouvain.be}
	\and
	Charles Pecheur
	\institute{
		ICTEAM Institute,\\
		Universit\'e catholique de Louvain,\\
		Louvain-la-Neuve, Belgium
	}
	\email{charles.pecheur@uclouvain.be}
}
\def\titlerunning{A Backward Approach for Model Checking Uniform Strategies for Constrained Reachability}
\def\authorrunning{S. Busard and C. Pecheur}
\begin{document}
\maketitle

\begin{abstract}
Since the introduction of Alternating-time Temporal Logic ($ATL$), many logics have been proposed to reason about different strategic capabilities of the agents of a system. In particular, some logics have been designed to reason about the uniform memoryless strategies of such agents. These strategies are the ones the agents can effectively play by only looking at what they observe from the current state.
$ATL_{ir}$ can be seen as the core logic to reason about such uniform strategies. Nevertheless, its model-checking problem is difficult---it requires a polynomial number of calls to an NP oracle---, and practical algorithms to solve it appeared only recently.

This paper proposes a technique for model checking uniform memoryless strategies. Existing techniques build the strategies from the states of interest---such as the initial states---through a forward traversal of the system. On the other hand, the proposed approach builds the winning strategies from the target states through a backward traversal, making sure that only uniform strategies are explored. Nevertheless, building the strategies from the ground up limits its applicability to constrained reachability objectives only.
This paper describes the approach in details and compares it experimentally with existing approaches implemented into a BDD-based framework. These experiments show that the technique is competitive on the cases it can handle.
\end{abstract}

\section{Introduction}

Alternating-time Temporal Logic ($ATL$) is one of the main logics to reason about strategies of the agents of a system~\cite{Alur-Henzinger-others-02}. Since its introduction 20 years ago, many extensions have been proposed, such as logics for reasoning about uniform strategies that agents with a partial view of the system can effectively play~\cite{Jamroga-Hoek-04}.
Unfortunately, extending $ATL$ for reasoning about uniform strategies with perfect recall yields an undecidable model-checking problem~\cite{Dima-Tiplea-11}. The problem can be made decidable by restricting it in several ways, such as considering hierarchical multi-player games~\cite{Peterson-Reif-others-02,Berthon-Maubert-others-17}, or restricting the agents to communicate publicly~\cite{Ramanujam-Simon-10,Belardinelli-Lomuscio-others-17}.

Nevertheless, these restrictions still yield very difficult model-checking problems (EXPTIME-complete and harder). On the other hand, restricting to uniform memoryless strategies---i.e., strategies that the agents can play by looking at what they observe from the current state---results in a $\Delta^P_2$-complete problem\footnote{A $\Delta^P_2$ problem requires a polynomial number of calls to an NP oracle.}~\cite{Jamroga-Dix-06}. In this context, $ATL_{ir}$~\cite{Schobbens-04} can be viewed as the minimal core logic that reasons about uniform memoryless strategies. It can be used, for instance, to reason about the strategies of multi-agent programs~\cite{Dastani-Jamroga-10}.

To illustrate the problems and techniques this paper discusses, we will use the example of a card game proposed by Jamroga and van der Hoek~\cite{Jamroga-Hoek-04}. The game is played with three cards $A$, $K$ and $Q$, between a player and a dealer. $A$ wins over $K$, $K$ wins over $Q$, and $Q$ wins over $A$. First, the dealer gives one card to the player, keeps one and leaves the last one on table, face down. Then the player can keep his card or swap it with the one on the table. Finally, the player wins if his card wins over the dealer's.

In this example, we can ask if there is a strategy for the player to win. $ATL$ considers that the player sees all the cards. In this case, he has a winning strategy as he can keep his card when he is already winning and swap it otherwise.
This semantics is counterintuitive as the player can choose different actions in situations he cannot distinguish---for instance, keeping the $A$ when the dealer has the $K$, and swapping it if the dealer has the $Q$.
On the other hand, $ATL_{ir}$ considers uniform memoryless strategies only. Under this semantics, the player has no winning strategy as he would need to swap his card when he has the $K$ and the dealer the $A$, while keeping it if the dealer has the $Q$. In this example, $ATL_{ir}$ provides a more natural framework to reason about the strategies of the player as he cannot observe the dealer's card.

While $ATL_{ir}$ has been studied extensively, symbolic algorithms to solve its model-checking problem appeared only recently~\cite{Busard-Pecheur-others-15, Busard-Pecheur-others-14, Pilecki-Bednarczyk-others-14, Huang-Meyden-14}.
The first solution proposed by Busard et al.\ enumerates and checks all uniform strategies of the agents to find a winning one~\cite{Busard-Pecheur-others-13, Busard-Pecheur-others-15}. It has been shown to be highly ineffective compared to other solutions~\cite{Busard-Pecheur-others-14}, so this paper does not consider it further.

The second approach proposed by Busard et al.\ is based on the idea of partial strategies, that is, strategies that are defined only for states that matter~\cite{Busard-Pecheur-others-14}. From a given subset of states of interest---such as the initial states---, we can compute the partial strategies that are needed to determine whether there exists a winning uniform strategy for a given objective. These partial strategies are built by alternating between computing the moves reached from the current partial strategy and splitting these new moves into uniform subsets. When the adequate partial strategies are generated, they can be checked for the objective with fixpoint computations. In the sequel, this solution is called the \emph{partial approach}.

To make the approach more efficient in practice, Busard et al.\ proposed two optimizations. First, as different partial strategies cover different overlapping subsets of states, sub-formulas are re-evaluated again and again, for each strategy. To avoid recomputing the truth value of sub-formulas, the results are cached.
The second optimization is early termination. It keeps track of the states of interest for which a winning strategy has already been found, and stops the process as soon as no states remain.

Pilecki et al.\ went further on the idea of partial strategies by showing that we do not need to determine a partial strategy in all states that matter before concluding~\cite{Pilecki-Bednarczyk-others-14}. During the process of discovering all these states that matter from the states of interest, we can check whether all extensions of the current partial strategy are winning or not, and stop if it is the case.

This idea can be improved further by also checking whether there exists a winning \emph{general} (not necessarily uniform) strategy extending the current one. If this is not the case, then there exists no such uniform strategy, and we can stop extending the current strategy and explore other choices.
Caching and early termination can also be applied. In the sequel, this solution is called the \emph{early approach}.

Finally, Huang and van der Meyden proposed to solve the model-checking problem by deriving, from the system under consideration, a new model where the uniform strategies of the agents are encoded into the derived states~\cite{Huang-Meyden-14}. Then we can compute the set of all winning uniform strategies by performing fixpoint computations on the derived model. In the sequel, this solution is called the \emph{symbolic approach}.

These approaches can be improved with pre-filtering, a technique that reduces the number of strategies to consider~\cite{Busard-Pecheur-others-13, Busard-Pecheur-others-15}. If some move does not belong to a winning \emph{general} strategy, then it cannot belong to a \emph{uniform} winning one. Furthermore, computing these losing moves can be done efficiently. Thus we can pre-compute these losing moves and ignore them when generating and checking strategies.

The partial and early approaches benefit from pre-filtering by ignoring losing moves when they build partial strategies. The symbolic approach ignores losing moves when encoding the uniform strategies in the states of the derived model, reducing their number.

Finally, the partial and early approaches can be implemented in a \emph{semi-symbolic} framework in which the strategies are represented with binary decision diagrams (BDDs~\cite{Bryant-86}), and checked symbolically using fixpoint computations. On the other hand, the symbolic approach fits a \emph{fully} symbolic framework as the derived model can be encoded with BDDs and directly checked with similar fixpoint computations.

The objective of this paper is to describe a new approach---the \emph{backward approach}---and to compare it with the existing ones.
The partial and early approaches enumerate the uniform BDD-encoded strategies through a forward traversal of the system, starting from the states of interest. The symbolic approach computes these winning uniform strategies through a fully symbolic backward traversal of the system. On the other hand, the proposed backward approach explicitly enumerates the BDD-encoded strategies through a \emph{backward} traversal from the target states. Unfortunately, this idea of computing the winning strategies from the target states is only applicable to constrained reachability objectives. These objectives deal with the existence of strategies that reach some particular states in a finite number of steps, potentially through some other particular states. For instance, winning the card game is a reachability objective.

The remainder of this paper is structured as follows. First, Section~\ref{section:background} reminds the syntax and semantics of $ATL_{ir}$. Section~\ref{section:backward-approach} describes the backward approach and Section~\ref{section:experimental-comparison} compares it with the existing symbolic approaches. Finally, Section~\ref{section:conclusion} concludes.

\section{Alternating-time Temporal Logic with Uniform Strategies}\label{section:background}

$ATL_{ir}$ formulas are composed of atomic propositions, the standard Boolean operators, and the $ATL$ strategic operators. More precisely, $ATL_{ir}$ formulas follow this grammar:
\begin{align*}
	\phi &::= true \mid p \mid \neg \phi \mid \phi \vee \phi \mid \CE{\Gamma}{\psi} \\
	\psi &::= \X{\phi} \mid \U{\phi}{\phi} \mid \W{\phi}{\phi}
\end{align*}
where $p$ is an atomic proposition of a set $AP$ and $\Gamma$ is a subset of a set of agents $Ag$.
The other standard Boolean operators ($\phi \wedge \phi$, $\phi \implies \phi$, $\phi \iff\phi$), and $ATL$ operators ($\CA{\Gamma}{\psi}$, $\G{\phi}$, $\F{\phi}$) can be defined in terms of these ones.

$ATL_{ir}$ formulas are interpreted over the states of \emph{imperfect information concurrent game structures} (iCGS). An iCGS is a structure $S = \langle Ag, Q, Q_0, Act, e, \delta, \sim, V\rangle$ such that
\begin{itemize}
	\item $Ag$ is a finite set of \emph{agents};
	\item $Q$ is a finite set of \emph{states};
	\item $Q_0 \subseteq Q$ is the set of \emph{initial} states;
	\item $Act$ is a finite set of \emph{actions}; a \emph{joint action} is a tuple $a \in Act^{Ag}$ of actions, one for each agent of $Ag$;
	\item $e: Ag \rightarrow (Q \rightarrow (2^{Act}\backslash\emptyset))$ defines, for each agent $ag$ and state $q$, the non-empty set of actions $ag$ can choose in $q$, that is, the actions \emph{enabled} in $q$; we write $e_{ag}$ for the function $e(ag)$ giving the set of actions $ag$ can choose in any state;
	\item $\delta: Q \times Act^{Ag} \nrightarrow Q$ is a partial deterministic \emph{transition function} defined for each state $q \in Q$ and each joint action enabled in $q$; we write $q \xrightarrow{a} q'$ for $\delta(q, a) = q'$;
	\item $\sim: Ag \rightarrow 2^{Q\times Q}$ defines a set of \emph{equivalence classes} representing the observability of agents; we write $\sim_{ag}$ for $\sim\!(ag)$ and we assume that each agent can choose his actions based on his own knowledge of the system, that is, $\forall q, q'\in Q, q\sim_{ag}q' \implies e_{ag}(q) = e_{ag}(q')$ for any agent $ag \in Ag$;
	\item $V: Q \rightarrow 2^{AP}$ is a function \emph{labeling} states with atomic propositions from a given set $AP$.
\end{itemize}

Given a set of agents $\Gamma \subseteq Ag$, we write $[Q']_\Gamma = \{q' \in Q \mid \exists q \in Q', \exists ag \in \Gamma, \text{ s.t. } q \sim_{ag} q'\}$ for the set of states indistinguishable by some agent $ag \in \Gamma$ from a state of $Q'$.

A joint action $a\in Act^{Ag}$ \emph{completes} an action $a_\Gamma \in Act^{\Gamma}$ for a set of agents $\Gamma$, written $a_\Gamma \sqsubseteq a$, if the action for each agent of $\Gamma$ in $a$ corresponds to the action of the same agent in $a_\Gamma$. Given a joint action $a \in Act^{Ag}$ and a set of agents $\Gamma \subseteq Ag$, we write $a(\Gamma)$ for the tuple of actions of agents of $\Gamma$ in $a$; when $\Gamma = \{ag\}$ is a singleton, we write $a(ag)$ instead of $a(\{ag\})$. The function $E: 2^{Ag} \rightarrow (Q \rightarrow 2^{Act^{Ag}})$ is defined as $E(\Gamma)(q) = \prod_{ag\in \Gamma} e_{ag}(q)$ and returns the set of actions for $\Gamma$ enabled in $q$; we write $E_{\Gamma}$ for $E(\Gamma)$. Finally, we call a \emph{$\Gamma$-move} (or a \emph{move} if $\Gamma$ is clear from the context) an element $\langle q, a_\Gamma\rangle \in Q\times Act^{\Gamma}$ such that $a_\Gamma \in E_{\Gamma}(q)$, that is, a pair composed of a state and an action for $\Gamma$ enabled in the state. We say that two $\Gamma$-moves $\langle q, a_\Gamma\rangle$ and $\langle q', a'_\Gamma\rangle$ are \emph{$\Gamma$-conflicting} if $\exists ag \in \Gamma \text{ s.t. } q\sim_{ag}q' \text{ and } a_\Gamma(ag) \neq a'_\Gamma(ag)$.
In other words, $\langle q, a_\Gamma\rangle$ and $\langle q', a'_\Gamma\rangle$ are $\Gamma$-conflicting if the states are indistinguishable for some agent $ag \in \Gamma$ and the proposed actions for $ag$ are different.
Furthermore, we say that a set of $\Gamma$-moves $M_\Gamma$ is $\Gamma$-conflicting if there exist two $\Gamma$-conflicting moves in $M_\Gamma$.

A \emph{path} in an iCGS $S$ is a sequence $\pi = q_0 \xrightarrow{a_1} q_1 \xrightarrow{a_2}...$ such that $\delta(q_d, a_{d+1}) = q_{d+1}$ for all $d \geq 0$. We write $\pi(d)$ for $q_d$, and $|\pi|$ for the number of states of $\pi$. If $\pi$ is infinite, $|\pi| = \omega$.
A \emph{memoryless strategy} for agent $ag$ is a function $f_{ag}: Q \rightarrow Act$ such that $\forall q\in Q, f_{ag}(q) \in e_{ag}(q)$. A (memoryless) \emph{uniform} strategy for agent $ag$ is a strategy $f_{ag} \text{ s.t. }\forall q, q' \in Q, q \sim_{ag} q' \implies f_{ag}(q) = f_{ag}(q')$. We call \emph{outcomes} of a strategy the infinite paths of the structure that are coherent with the strategy. More precisely, the outcomes of a strategy $f_{ag}$ for agent $ag$ from a state $q$ are defined as
\begin{align}
	out(f_{ag}, q) = \{\pi = q_0 \xrightarrow{a_1} q_1 \xrightarrow{a_2}... \mid q_0 = q \wedge \forall d\in \mathbb{N}, f_{ag}(q_d) \sqsubseteq a_{d+1}\}.
\end{align}
A (uniform) strategy for a group of agents $\Gamma \subseteq Ag$ is a tuple of (uniform) strategies, one for each agent of $\Gamma$. The outcomes of a strategy $f_\Gamma$ for a group of agents $\Gamma$ from a state $q$ are defined as
\begin{align}
	out(f_\Gamma, q) = \bigcap_{f_{ag}\in f_\Gamma} out(f_{ag}, q).
\end{align}
These outcomes are the paths that are coherent with every strategy of the set $f_\Gamma$.
Finally, the outcomes function $out$ is lifted for any subset of $\Gamma$-moves $M_\Gamma$ as follows:
\begin{align}
	out(M_\Gamma, q) = \{ \pi = q_0 \xrightarrow{a_1} q_1 \xrightarrow{a_2}... \mid q_0 = q \wedge \forall d, 0 \leq d < |\pi|-1,
	\exists \langle q', a'_\Gamma\rangle \in M_\Gamma \text{ s.t. }
	q' = q_d \wedge a'_\Gamma \sqsubseteq a_{d+1}\},
\end{align}
that is, $out(M_\Gamma, q)$ is the set of (finite or infinite) paths that follow some actions for $\Gamma$ proposed by $M_\Gamma$.

In the sequel, we mainly speak about uniform strategies and call them strategies. When speaking about strategies that are not necessarily uniform, we speak about \emph{general} strategies. A strategy $f_\Gamma$ can be represented as the set of $\Gamma$-moves $\{ \langle q, a_\Gamma \rangle \in Q\times Act^{\Gamma} \mid a_\Gamma = f_\Gamma(q)\}$, that is, the set of moves such that the actions are the ones specified by the strategy. In the sequel, the notation $f_\Gamma$ is interchangeably used for a set of $\Gamma$-moves and the strategy they represent.
Furthermore, we say that a set of $\Gamma$-moves $M_\Gamma$ \emph{covers} a set of states $Q' \subseteq Q$ if $\forall q \in Q', \exists \langle q',a'_\Gamma\rangle \in M_\Gamma \text{ s.t. } q' = q$. In other words $M_\Gamma$ covers $Q'$ if $M_\Gamma$ proposes an action for all states of $Q'$. We write $M_\Gamma |_Q$ for the set of states $M_\Gamma$ covers.
We also interchangeably write $E_\Gamma$ for the original function taking a state $q$ and returning the set of actions $\Gamma$ can play in $q$, and for the set of $\Gamma$-moves it represents, that is, the set $\{\langle q, a_\Gamma \rangle \in Q \times Act^{\Gamma} \mid a_\Gamma \in E_\Gamma(q)\}$. Finally, the function $Moves_{\Gamma}(Q') = \{\langle q', a'_\Gamma\rangle \in E_\Gamma \mid q' \in Q'\}$ returns the set of $\Gamma$-moves enabled in states of $Q'$.

The semantics of $ATL_{ir}$ is defined over states of an iCGS $S$ by the relation $S, q \models \phi$. $S$ is omitted when clear from the context. This relation meets the standard semantics for Boolean operators. For strategic operators, the $q \models \phi$ relation is defined as
\begin{align*}
	q \models \CE{\Gamma}{\psi} \Leftrightarrow \exists\text{ a \textit{uniform strategy} $f_\Gamma$ for $\Gamma$ s.t. }\forall ag \in\Gamma,\forall q' \sim_{ag} q, \forall\text{ paths } \pi \in out(f_\Gamma, q'),\pi \models \psi.
\end{align*}
The relation $\pi \models \psi$ over paths $\pi$ of the structure $S$ is defined in the standard way as
\begin{align*}
	&\pi \models \X{\phi} &\ \Leftrightarrow\ &\pi(1) \models \phi, \\
	&\pi \models \U{\phi_1}{\phi_2} &\ \Leftrightarrow\ &\exists d \geq 0 \text{ s.t. } \pi(d) \models \phi_2 \text{ and }\forall e < d, \pi(e) \models \phi_1, \\
	&\pi \models \W{\phi_1}{\phi_2} &\ \Leftrightarrow\ &\exists d \geq 0 \text{ s.t. } \pi(d) \models \phi_2\text{ and } \forall e < d, \pi(e) \models\phi_1,\text{ or } \forall d \geq 0, \pi(d)\models \phi_1.
\end{align*}
We write $S \models \phi$ if all initial states of $S$ satisfy $\phi$, that is, if $\forall q \in Q_0, S, q\models \phi$.
Intuitively, this semantics says that $q$ satisfies $\CE{\Gamma}{\psi}$ if agents in $\Gamma$ have a collective strategy such that, whatever the actions of the other agents are, the objective $\psi$ is satisfied by all the resulting paths from all indistinguishable states.

\section{The Backward Approach}\label{section:backward-approach}

The main idea of the \emph{backward} approach is to generate the winning strategies through a backward exploration of the system.
For instance, let us consider the card game.
Because the player does not see the card on the table nor the card of the dealer before making a decision, he has no uniform strategy to win the game.
To check whether there exists a strategy to win the game---that is, whether $\CEF{player}{win}$ is satisfied---, we can start by looking at the states in which the player already wins the game, and look at the non-conflicting moves that can reach these states. By iterating this procedure, we can explore the parts of the uniform strategies that surely reach the winning states.

\begin{figure}[!ht]
	\centering
	\scalebox{0.8}{
	\includegraphics{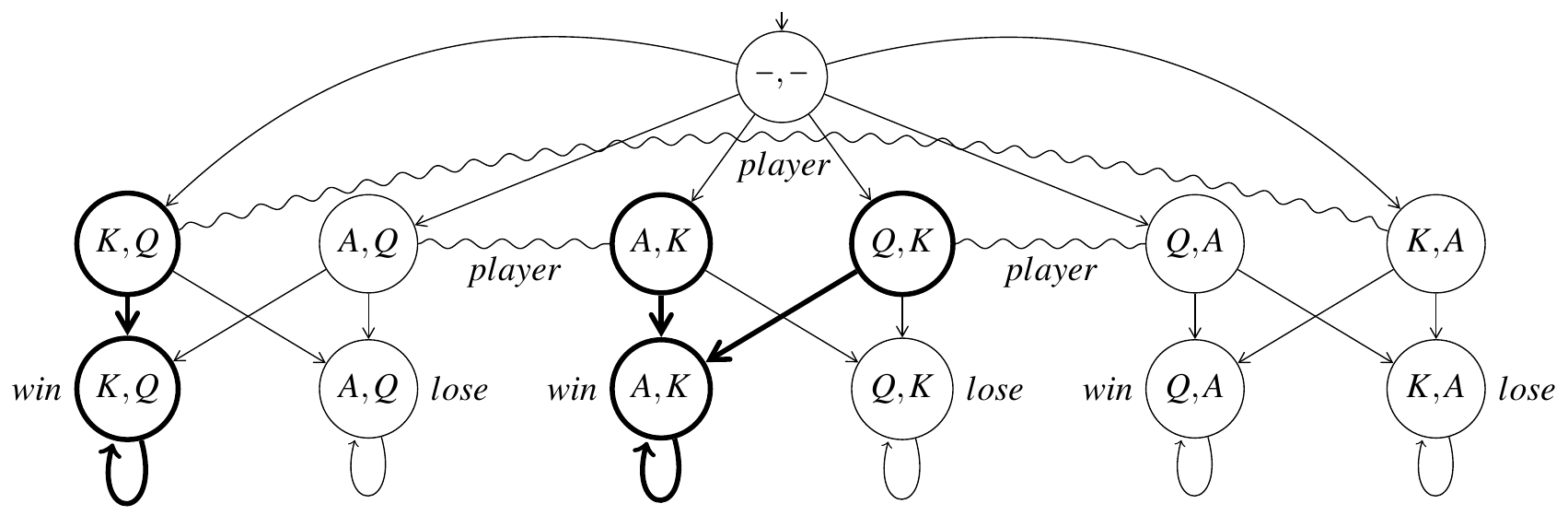}
	}
	\caption{The graph of the card game. States are labelled with $C_1, C_2$ when the player has card $C_1$ and the dealer has $C_2$. Arrows are temporal transitions, actions of the two players are easily inferred. The wavy edges link states that are indistinguishable by the player. In bold, the winning part of a uniform strategy that chooses to swap the card when the player has the $Q$ and to keep it otherwise.}
	\label{figure:simple-card-game-backward}
\end{figure}

Figure~\ref{figure:simple-card-game-backward} shows the graph of the card game with the winning parts of a uniform strategy in bold. This strategy chooses to swap the card when the player has $Q$ and to keep it otherwise. This set of non-conflicting moves cannot be extended with non-conflicting moves that would surely reach the set. Thus no uniform strategy that makes these choices is winning for the initial state, because the initial state has no move in the set. There exist other subsets of moves that make the player reach the state in which he wins, but they are not winning in the initial state either, so the player has no uniform strategy to win.

The backward approach uses this idea of generating the winning parts of the uniform strategies from the target states.
Nevertheless, it cannot handle greatest fixpoint-based objectives because, in this case, we cannot build the winning strategies from the ground up. The approach thus cannot handle $\CEW$ and $\CEG$ objectives.
More precisely, it can handle all $ATL_{ir}$ formulas generated through the grammar
\begin{align*}
	\phi &::= true \mid p \mid \neg \phi \mid \phi \vee \phi \mid \CE{\Gamma}{\psi} \\
	\psi &::= \X{\phi} \mid \U{\phi}{\phi}
\end{align*}
In particular, it can handle the $\CAG$ and $\CAW$ strategic operators (through negation), but not the $\CEG$, $\CEW$, $\CAU$, and $\CAF$ ones.

The approach (see Algorithm~\ref{algo:eval-backward-irF}) uses the $filter_{\CEU}$ algorithm, and the $Pre^M$, $Compatible^M$, $SplitMax$, $SplitNonEmpty$, and $Post$ functions.
The $filter_{\CEU}$ algorithm is based on the $Pre_{\CE{\Gamma}}$ function defined as
\begin{align}
	Pre_{\CE{\Gamma}}(Q') = \{q \in Q \mid \exists \langle q, a_\Gamma\rangle \in E_\Gamma \text{ s.t. } \forall a \in E_{Ag}(q), a_\Gamma \sqsubseteq a \implies \delta(q, a) \in Q' \}.
\label{eq:pre-ce-gamma}
\end{align}
Intuitively, $Pre_{\CE{\Gamma}}(Q')$ returns the states $q \in Q$ such that there exists an action for $\Gamma$ in $q$ that surely leads to a state of $Q'$ in one step.
Then $filter_{\CEU}$ is defined as
\begin{align}
	filter_{\CEU}(Q_1, Q_2) = \mu Z. Q_2 \cup (Q_1 \cap Pre_{\CE{\Gamma}}(Z)).
\end{align}
It takes a set of agents $\Gamma \subseteq Ag$ and two sets of states $Q_1, Q_2 \subseteq Q$, and returns the states $q$ such that there is a \emph{general} strategy that forces to reach $Q_2$ through $Q_1$ from $q$.
$Pre^{M}_{\CE{\Gamma}}$ is a variant of $Pre_{\CE{\Gamma}}$ defined as
\begin{align}
	Pre^{M}_{\CE{\Gamma}}(M'_\Gamma) = \{\langle q, a_\Gamma\rangle \in E_\Gamma \mid \forall a \in E_{Ag}(q), a_\Gamma \sqsubseteq a \implies \delta(q, a) \in M'_\Gamma|_Q \}.
\end{align}
It takes a set of $\Gamma$-moves $M'_\Gamma$ and returns the set of $\Gamma$-moves  reaching only states of moves of $M'_\Gamma$.
The $Compatible^M$ function is defined as
\begin{align}
	Compatible^M(M'_\Gamma, M_\Gamma) = \{\langle q', a'_\Gamma\rangle \in M'_\Gamma \mid\ \not\exists \langle q, a_\Gamma \rangle \in M_\Gamma, ag \in \Gamma\text{ s.t. } q \sim_{ag} q' \wedge a_\Gamma(ag) \neq a'_\Gamma(ag)\}.
\end{align}
It takes two subsets of $\Gamma$-moves and returns the moves of $M'_\Gamma$ that are compatible with moves of $M_\Gamma$.

The $SplitMax$ function takes a set of agents $\Gamma \subseteq Ag$ and a set of $\Gamma$-moves $M_\Gamma$ and returns all the largest subsets of non-$\Gamma$-conflicting moves of $M_\Gamma$~\cite{Busard-Pecheur-others-14}. The $SplitNonEmpty$ function takes the same arguments and returns the set of non-empty subsets of non-$\Gamma$-conflicting equivalence classes of moves of $M_\Gamma$.
Each such subset $M'_\Gamma$ represents (a part of) a uniform strategy. Indeed, $M'_\Gamma$ proposes joint actions for $\Gamma$ such that, for any agent $ag \in \Gamma$, for two states indistinguishable by $ag$, $M'_\Gamma$ gives the same action for $ag$.

Both functions are based on the $SplitAll$ algorithm (see Algorithm~\ref{algo:split-gamma}), that is based on the $SplitAgent$ algorithm (see Algorithm~\ref{algo:split-agent}). The latter takes a set of moves $M_\Gamma$ for $\Gamma$, an agent $ag \in \Gamma$ and a boolean $max$, and returns the set of subsets of non-$ag$-conflicting equivalence classes of moves of $M_\Gamma$, restricting this set to the largest subsets if $max$ is $true$. It goes through all equivalence classes of $M_\Gamma$ for $ag$ and split them into non-$ag$-conflicting subsets.
The $SplitAll$ algorithm computes the set of subsets of non-$\Gamma$-conflicting equivalence classes of moves of $M_\Gamma$. It uses the $SplitAgent$ algorithm to split $M_\Gamma$ for each agent of $\Gamma$.

\begin{algorithm}[!ht]
	\DontPrintSemicolon
	\LinesNumberedHidden
	\KwData{$ag \in \Gamma$ an agent of $\Gamma$,  $\Gamma \subseteq Ag$ a group of agents, $M_\Gamma \subseteq E_\Gamma$ a set of $\Gamma$-moves, $max$ a boolean.}
	\KwResult{The set of subsets of non-$ag$-conflicting equivalence classes of moves of $M_\Gamma$. If $max$ is $true$, then only the largest ones are returned.}
	\BlankLine
	$\langle q, a_\Gamma \rangle = $\textbf{ pick} one element in $M_\Gamma$\;
	$equivalent = \{ \langle q', a'_\Gamma \rangle \in M_\Gamma \mid q' \sim_{ag} q \}$\;
	$actions = \{ a_{ag} \in Act  \mid  \exists \langle q', a'_\Gamma \rangle \in equivalent \text{ s.t. } a'_\Gamma(ag) = a_{ag}\}$\;
	$ncsubsets = SplitAgent(ag, \Gamma, M_\Gamma \backslash equivalent)$\;
	$subsets = \{\}$\;
	\For{$a_{ag} \in actions$}{
		$equivsubset = \{ \langle q', a'_\Gamma \rangle \in equivalent  \mid  a'_\Gamma(ag) = a_{ag} \}$\;
		$subsets = subsets \cup \left\{equivsubset \cup ncsubset  \mid  ncsubset \in ncsubsets\right\}$\;
	}
	\lIf{$\neg max$}{$subsets = subsets \cup ncsubsets$}
	\Return{$subsets$\;}
	\caption{$SplitAgent(ag, \Gamma, M_\Gamma, max)$}
	\label{algo:split-agent}
\end{algorithm}

\begin{algorithm}[!ht]
	\DontPrintSemicolon
	\LinesNumberedHidden
	\KwData{$\Gamma \subseteq Ag$ a group of agents,  $M_\Gamma \subseteq E_\Gamma$  a set of moves, $max$ a boolean.}
	\KwResult{The set of subsets of non-$\Gamma$-conflicting equivalence classes of moves of $M_\Gamma$. If $max$ is $true$, then only the largest ones are returned.}
	\BlankLine
	$subsets = \{M_\Gamma\}$\;
	\For{$ag \in \Gamma$}{
		$subsets' = \{\}$\;
		\lFor{$subset \in subsets$}{
			$subsets' = subsets' \cup SplitAgent(ag, \Gamma, subset, max)$
		}
		$subsets = subsets'$\;
	}
	\Return{$subsets$}\;
	\caption{$SplitAll(\Gamma, M_\Gamma, max)$}
	\label{algo:split-gamma}
\end{algorithm}

The $SplitNonEmpty$ function is then defined as
\begin{align}
	SplitNonEmpty(\Gamma, M_\Gamma) = \{M'_\Gamma \in SplitAll(\Gamma, M_\Gamma, false) \mid M'_\Gamma \supset \emptyset\},
\end{align}
and the $SplitMax$ one as
\begin{align}
	SplitMax(\Gamma, M_\Gamma) = SplitAll(\Gamma, M_\Gamma, true).
\end{align}

The $Post$ function takes a set of states $Q' \subseteq Q$ and returns the successor states of states of $Q'$. Formally,
\begin{align}
	Post(Q') = \{q \in Q \mid \exists q' \in Q', \exists a' \in E_{Ag}(q')\text{ s.t. } \delta(q', a') = q\}.
\end{align}

Finally, Algorithm~\ref{algo:eval-backward-irF} uses the $eval_{\CEU}$ algorithm (see Algorithm~\ref{algo:eval-backward-ceu-irF}) to compute the states for which there exists a strategy to win a $\CEU$ objective.

Let $Q_1, Q_2\subseteq Q$ be two subsets of states. We say that a non-$\Gamma$-conflicting subset of $\Gamma$-moves $M_\Gamma$ \emph{enforces to reach $Q_2$ through $Q_1$} if $Q_2 \subseteq M_\Gamma|_Q$, and for all states $q \in M_\Gamma|_Q$, for all paths $\pi \in out(M_\Gamma, q)$, $\pi$ is finite and $\pi(|\pi|) \in Q_2 \wedge \forall d, 0 \leq d < |\pi|, \pi(d) \in Q_1 \backslash Q_2$, or $\pi$ is infinite and there is a finite prefix of $\pi$ that satisfies the conditions above.
In other words, $M_\Gamma$ enforces to reach $Q_2$ through $Q_1$ if all the paths enforced by $M_\Gamma$ reach a state of $Q_2$ through states of $Q_1\backslash Q_2$.

Given two formulas $\phi_1$ and $\phi_2$, there exists a strategy $f_\Gamma$ such that all outcomes from some state $q$ satisfy $\U{\phi_1}{\phi_2}$ iff there exists a subset of moves $M'_\Gamma$ containing a move for $q$ that enforces to reach states satisfying $\phi_2$ through states satisfying $\phi_1$. The $eval_{\CEU}$ algorithm uses this property to compute the states for which there exists a winning strategy for a $\CEU$ objective.

More precisely, it takes as arguments a subset $Q' \subseteq Q$ such that $Q' = [Q']_\Gamma$, $M_\Gamma \subseteq E_\Gamma$ a non-conflicting set of moves, and two subsets of states $Q_1, Q_2 \subseteq Q$ such that $M_\Gamma$ \emph{enforces to reach $Q_2$ through $Q_1$}. From these arguments, it computes the set of states $q \in Q'$ such that there exists a uniform strategy $f'_\Gamma$ that shares the same choices as $M_\Gamma$ and such that all outcomes of $f'_\Gamma$ from all states indistinguishable from $q$ reach a state of $Q_2$ through states of $Q_1$.

To compute this set of states, $eval_{\CEU}$ first computes some states for which there surely cannot exist a winning \emph{general} strategy (in $lose$) and for which there exists a winning uniform strategy (in $win$). If $lose$ and $win$ cover all states of interest $Q'$, then the job is done. Otherwise, it computes the moves $compatible$ from states of $Q_1$ that can surely reach $M_\Gamma$ and are compatible with it, and recursively calls itself with $M_\Gamma$ extended with the non-empty non-conflicting subsets of $compatible$, accumulating the results in $win$.
It uses an additional $exclude$ parameter to exclude from the following steps the moves of $new\_moves$ it ignored. This feature is not necessary, but makes the algorithm more efficient as it has not to consider the excluded moves again and again.

\begin{algorithm}[!ht]
	\DontPrintSemicolon
	\LinesNumberedHidden
	\KwData{$Q' \subseteq Q$ a subset of states such that $Q' = [Q']_\Gamma$, $M_\Gamma \subseteq E_\Gamma$ a non-$\Gamma$-conflicting set of $\Gamma$-moves, $Q_1, Q_2 \subseteq Q$ two subsets of states such that $M_\Gamma$ enforces to reach $Q_2$ through $Q_1$, $exclude \subseteq E_\Gamma$ a subset of moves such that $exclude \cap M_\Gamma = \emptyset$.}
	\KwResult{The states $q \in Q'$ such that there exists a uniform strategy $f'_\Gamma \supseteq M_\Gamma$ such that $f'_\Gamma \cap exclude = \emptyset$ and all  outcomes of $f'_\Gamma$ from all states indistinguishable from $q$ reach a state of $Q_2$ through states of $Q_1$.}
	\BlankLine
	$notlose = filter_{\CEU}(Q_1, M_\Gamma|_Q)$\;
	$lose = \{q \in Q' \mid \exists ag \in \Gamma \text{ s.t. } \exists q' \in Q \text{ s.t. } q' \sim_{ag} q \wedge q' \not\in notlose\}$\;
	$win = \{q \in Q' \mid \forall ag \in \Gamma, \forall q' \in Q, q' \sim_{ag} q \implies q' \in M_\Gamma|_Q\}$\;
	\lIf{$Q' \backslash (lose \cup win) = \emptyset$}{
		\Return{$win$}
	}
	\Else{
		$Q' = Q' \backslash (lose \cup win)$\;
		$new\_moves = (Pre^M_{\CE{\Gamma}}(M_\Gamma) \cap Moves_{\Gamma}(Q_1)) \backslash M_\Gamma$\;
		$new\_moves = new\_moves \backslash exclude$\;
		$compatible = Compatible^M(new\_moves, M_\Gamma)$\;
		\lIf{$compatible = \emptyset$}{
			\Return{$win$}
		}
		\Else{
			\For{$M'_\Gamma \in SplitNonEmpty(\Gamma, compatible)$}{
				$win = win \cup eval_{\CEU}(Q', M_\Gamma \cup M'_\Gamma, Q_1, Q_2, exclude \cup (new\_moves \backslash M'_\Gamma)))$\;
				$Q' = Q' \backslash win$\;
				\lIf{$Q' = \emptyset$}{
					\Return{$win$}
				}
			}
			\Return{$win$}\;
		}
	}
	\caption{$eval_{\CEU}(Q', M_\Gamma, Q_1, Q_2, exclude)$}
	\label{algo:eval-backward-ceu-irF}
\end{algorithm}

The $eval$ algorithm (see Algorithm~\ref{algo:eval-backward-irF}) can handle $\CEX$ and $\CEU$ formulas. For $\CEX$, it recursively computes the states of $S$ satisfying the sub-formula $\phi'$ and then computes the states for which there exists a move for all indistinguishable states. More precisely, it splits the set of moves that $\Gamma$ can use to enforce to reach the states satisfying $\phi'$ into non-conflicting greatest subsets. There exists a strategy that wins the objective in $q$ iff there exists an action that enforces to reach states of $Q'''$ in one step in all states indistinguishable from $q$, and that is exactly what is computed by the algorithm and accumulated in $sat$.

For $\CEU$, it recursively computes the states of $S$ satisfying the sub-formulas $\phi_1$ and $\phi_2$. Then is uses the $eval_{\CEU}$ algorithm with the greatest non-conflicting subsets of the moves of the states satisfying $\phi_2$ to accumulate in $sat$ the states $q \in Q''$ such that there exists a strategy to win the objective in all states indistinguishable from $q$. The actions chosen by these moves for the states satisfying $\phi_2$ are not significant for the winning strategies as the states already satisfy $\phi_2$, but are necessary for the $eval_{\CEU}$ algorithm to work properly.

\begin{algorithm}[!ht]
	\DontPrintSemicolon
	\LinesNumberedHidden
	\KwData{$S$ an iCGS, $Q' \subseteq Q$ a subset of states, $\phi$ an $ATL_{ir}$ formula.}
	\KwResult{The states of $Q'$ satisfying $\phi$.}
	\BlankLine
		\Case{$\phi \in \{\CEX{\Gamma}{\phi'}, \CEU{\Gamma}{\phi_1}{\phi_2}\}$}{
   		$Q'' = [Q']_\Gamma$;
   		$sat = \emptyset$\;
   		\Case{$\phi = \CEX{\Gamma}{\phi'}$}{
   			$Q''' = eval(S, Post([Q'']^E_\Gamma), \phi')$\;
   			\For{$M_\Gamma \in SplitMax(\Gamma, Pre^{M}_{\CE{\Gamma}}(Moves_{\Gamma}(Q''')))$}{
   				$sat = sat \cup \{q \in Q'' \mid \forall ag \in \Gamma, \forall q'\in Q, q' \sim_{ag} q \implies q' \in M_\Gamma|_Q\}$\;
   				$Q'' = Q'' \backslash sat$\;
   				\lIf{$Q'' = \emptyset$}{
   					\Return{$sat \cap Q'$}
   				}
   			}
   			\Return{$sat \cap Q'$}
   		}
   		\Case{$\phi = \CEU{\Gamma}{\phi_1}{\phi_2}$}{
   			$Q_1 = eval(S, Q, \phi_1)$;
   			$Q_2 = eval(S, Q, \phi_2)$\;
				$sat = \{q \in Q'' \mid\forall ag \in \Gamma, \forall q' \in Q, q' \sim_{ag} q \implies q' \in Q_2\}$\;
   			\lIf{$sat = Q''$}{
   				\Return{$sat \cap Q'$}
   			}
   			$Q'' = Q'' \backslash sat$\;
   			\For{$M_\Gamma \in SplitMax(\Gamma, Moves_\Gamma(Q_2))$}{
   				$sat = sat \cup eval_{\CEU}(Q'', M_\Gamma, Q_1, Q_2, \emptyset)$\;
   				$Q'' = Q'' \backslash sat$\;
   				\lIf{$Q'' = \emptyset$}{
   					\Return{$sat \cap Q'$}
   				}
   			}
   			\Return{$sat \cap Q'$}
   		}
		}
		\textit{// $\CEW{\Gamma}{\phi_1}{\phi_2}$ is not supported}\;
		\textit{// Boolean cases are handled in the standard way}
	\caption{$eval(S, Q', \phi)$}
	\label{algo:eval-backward-irF}
\end{algorithm}

While existing approaches such as the partial, early and symbolic ones can reduce the number of strategies by pre-filtering losing moves, the backward approach does not benefit from this idea. The approach already explores winning choices only, ignoring the losing ones.

\section{Experimental Comparison}\label{section:experimental-comparison}

This section experimentally compares the existing approaches for model checking uniform strategies to the backward one. It first describes the model and formulas the experiments are based on, and then presents the experimental results themselves.
All the approaches, including the backward one, have been implemented with BDDs thanks to PyNuSMV~\cite{Busard-Pecheur-13}, a Python framework based on the state-of-the-art model checker NuSMV~\cite{Cimatti-Clarke-others-02}.
These implementations are shipped with PyNuSMV. Explanations on how to reproduce the experiments can be found at \url{http://lvl.info.ucl.ac.be/GandALF2017}.

\subsection{Model and Properties}

The model used for the experiments is the model of the three castles already used by Pilecki et al.\ for their own experiments in~\cite{Pilecki-Bednarczyk-others-14}. It is composed of three castles with their corresponding health points ranging from $0$ to $3$, $0$ health points meaning that the castle is defeated. Each castle is defended by a set of workers. At each turn, a worker can attack another castle, defend his own castle or do nothing, but a worker cannot defend her castle twice in a row. The number of damage points a castle receives is the number of attackers against this castle minus the number of defenders of this castle, if this number is greater than $0$. The health points of the castles are not reset at each turn, thus the game is played in several turns. Finally, the workers only observe whether they can defend their castle or not, and, for each castle, whether it is defeated or not. They also distinguish the initial state from the others to be able to reason about the strategies they have in this initial state. The model is parametrized with the number of workers of each castle.

The depth of the model---that is, the number of steps needed to reach all the reachable states from the initial one---does not change with the number of workers since it depends only on the health points of the castles. An exception is when there is one worker in each castle. In this case, the depth is a bit higher because there are too few workers to ensure to quickly reach a final state. The partial, early and backward approaches really depend on this depth since it dictates how far the adequate partial strategies are.

We are interested in two formulas. The first one is $\phi_1 = \CEF{\mathit{Castle_1}, \mathit{Castle_2}}{\mathit{Castle_3\ defeated}}$, where $\mathit{Castle_i}$ groups the workers of the $i$th castle and $\mathit{Castle_3\ defeated}$ is true in all states in which the third castle has $0$ health points. This formula is true in all tested models, but is not true in general. If the third castle has enough workers, they are able to defend the castle and prevent the other workers to damage it. More precisely, if the third castle has more workers than the addition of the two others, the formula is false, even if the workers have perfect information. The tested models always have enough workers in the first two castles to make the formula satisfied.

The second formula is $\phi_2 = \CEF{\mathit{Worker_1}, \mathit{Worker_2}}{\mathit{all\ defeated}}$, where $\mathit{Worker_1}$ (resp. $\mathit{Worker_2}$) is a worker of the first castle (resp. second castle), and $\mathit{all\ defeated}$ is true in the states where all castles have $0$ health points. This formula is false in all tested models because, even if they can defeat the third castle, the workers have not enough information to ensure that the other two castles will be defeated at the same time. Indeed, they do not observe the remaining health points of the castles, and cannot attack their own castle (at any time) or the remaining one when their own is defeated.

\subsection{Experimental Results}

The two formulas have been checked using the approaches on models of increasing size. This section presents and compares the results. All the experiments have been performed on a MacBook Pro with a $2.6$GHz processor and $16$GB RAM, under a time limit of $1800$ seconds. This limit is indicated by a horizontal line in the graphs, and data points reaching it are depicted above the line.
Each data point is the average of $20$ runs; the observed variability was very low for all measurements. These experiments usually consumed less than $1$GB of memory, but some consumed up to several GBs. They nevertheless never consumed all the available memory.
For each approach, observations are given, then the differences of performances are explained based on these observations. In the sequel, variants of the approaches with pre-filtering are named by adding \emph{/filt} (e.g., the early approach with pre-filtering is named \emph{Early/filt}).

The Python implementation used for these experiments is a prototype showing the applicability of the approaches. It would not compete with dedicated tools performing the same kind of tasks. These experiments are not meant to show the absolute performances of the implementation but the relative gain of the different approaches.

\subsubsection{$\phi_1 = \CEF{\mathit{Castle_1}, \mathit{Castle_2}}{\mathit{Castle_3\ defeated}}$}

\begin{figure}[!ht]
	\centering
	\includegraphics{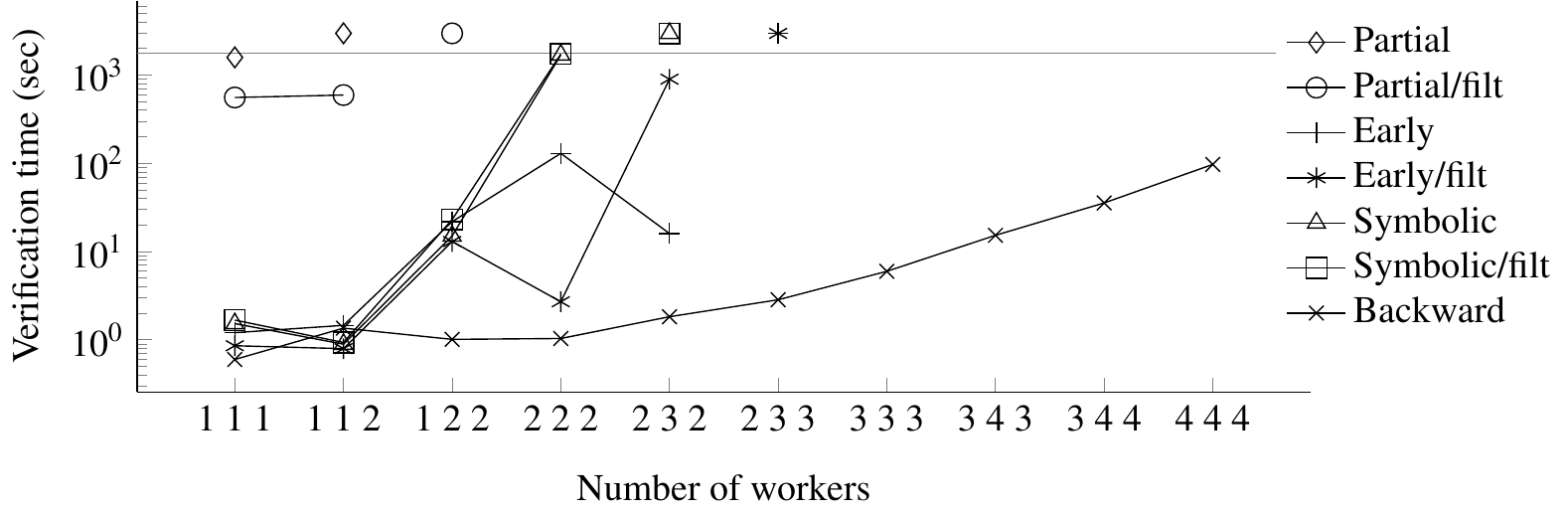}
	\caption{Evolution of the verification time for the formula $\CEF{\mathit{Castle_1}, \mathit{Castle_2}}{\mathit{Castle_3\ defeated}}$.}
	\label{figure:castles-phi1}
\end{figure}

Figure~\ref{figure:castles-phi1} shows the evolution of verification time of the seven approaches for checking the formula $\phi_1$ on the model of the castles. The size of the model (Number of workers) is given as a triplet $1\ 2\ 3$, meaning that the first castle is defended by one worker, the second one by two, and the third one by three workers.

Depending on the size of the model, pre-filtering removes from $18\%$ ($1\ 1\ 1$ case) to $77\%$ ($1\ 1\ 2$ case) of the moves. In the first case, the first two castles have more power than the third one and can easily win if they have perfect information. In the second case, the power of the first two castles is similar to the power of the third castle, and thus cannot easily win even with perfect information. For the other sizes, the gain is between these two bounds.

The Partial approach succeeds in finding a winning strategy within $30$ minutes for the $1\ 1\ 1$ case. Nevertheless, for the $1\ 1\ 2$ case, it cannot find a winning one. In this case, there is the same number of strategies, but it is more costly to check each strategy as the model is bigger.
On the other hand, the Partial/filt approach benefits from pre-filtering and finds a winning strategy more quickly than the Partial one. Nevertheless, it fails at finding a winning one in the $1\ 2\ 2$ case.

The Early approach needs to reach up to half the depth of the model to determine the strategies to be losing. This allows the approach to find a winning strategy easily.
The number of strategies increases with the number of workers to consider, as well as the time needed to check larger models. On the $2\ 3\ 2$ case, it finds a winning strategy more quickly because it makes the right choices earlier.

In the $2\ 2\ 2$ case, the Early/filt approach very quickly finds a winning strategy. It really benefits from pre-filtering and finds a good strategy after a few steps.
In the $2\ 3\ 2$ case, it needs to consider many more strategies before finding a good one.

The symbolic approaches have to encode and check all strategies at the same time. As the number of workers increases, there are more and more strategies for the group.

The Backward approach starts from the states in which Castle 3 is defeated and explores the moves that surely reach them. It needs to extend the strategies with moves that are as far as half the depth of the model to determine whether a strategy is losing or not in the initial state. This is especially true in the smaller models in which the workers of the first two castles have power comparable to those of the third castle. For larger models, the first two castles workers have more power than those of the third castle, and the approach needs only one or two steps, and no backtracking, to find a winning strategy. The increase of time is simply due to the fact that the model becomes larger and larger, and evaluating a single strategy---with the $filter_{\CEU}$ algorithm---takes more and more time.

\paragraph{Comparison}
The number of adequate partial strategies is large, and the partial approaches quickly fail to find a winning one.
The symbolic approaches are better. Nevertheless, pre-filtering does not benefit to the Symbolic/filt approach because all equivalence classes are still present and all actions are still possible in each of them, thus both approaches do the same work.

The early approaches are even better in the present scenario because they can quickly determine that a partial strategy and all its extensions cannot be winning. The Early/filt approach really benefits from pre-filtering and drastically reduces the number of strategies it checks for the larger models. The two approaches show some irregularities in performances because they sometimes make the right choices of actions, and sometimes not.

The backward approach is the best in this scenario because it concentrates on the strategies that can effectively reach the target states. It does not need to backtrack a lot before finding a winning strategy in the initial state.

\subsubsection{$\phi_2 = \CEF{\mathit{Worker_1}, \mathit{Worker_2}}{\mathit{all\ defeated}}$}

\begin{figure}[!ht]
	\centering
	\includegraphics{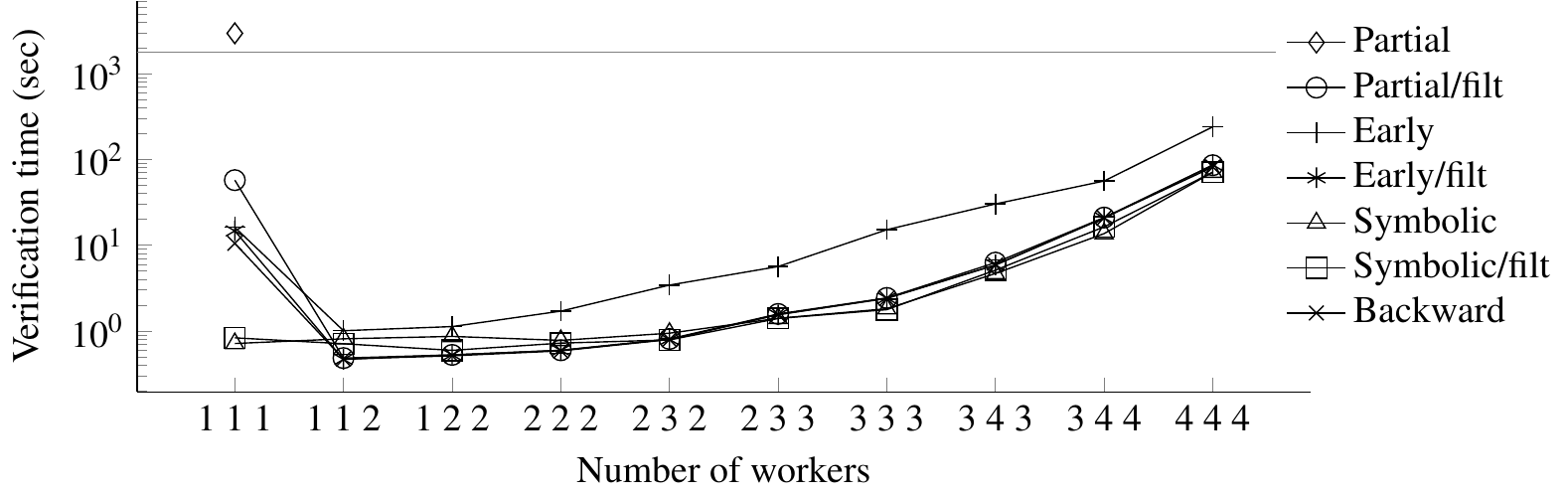}
	\caption{Evolution of the verification time for the formula $\CEF{\mathit{Worker_1}, \mathit{Worker_2}}{\mathit{all\ defeated}}$.}
	\label{figure:castles-phi2}
\end{figure}

Figure~\ref{figure:castles-phi2} shows the evolution of verification time of the seven approaches for checking the formula $\phi_2$. This formula is false for all checked sizes.

A major difference between the $1\ 1\ 1$ case and the others is that, in the former case, the two workers have a strategy to achieve their goal when they have perfect information, while it is not the case for the greater sizes. Thus, pre-filtering, in the cases of larger models, allows the Partial/filt and Early/filt approaches to directly determine that the formula is false, without checking any strategy.

The Partial approach reaches the timeout even for the smallest model size, failing at checking all $\approx 6.9\times 10^9$ adequate partial strategies.
On the other hand, pre-filtering drastically reduces the number of moves to consider for the $1\ 1\ 1$ case, and thus the number of strategies the Partial/filt approach needs to check before stating that the formula is false.

For the $1\ 1\ 1$ case, the Early approach also needs to reach about half way from the initial state to determine strategies to be losing, as for the previous formula. It so checks all strategies more easily.
For the other cases, the approach only needs to check the $16$ initial actions of the two workers to conclude that there can be no winning strategy.
For the $1\ 1\ 2$ case and beyond, the Early/filt approach does not check any strategy since pre-filtering directly determines that there can be no winning strategy.

The Symbolic approach behaves in the same way for all model sizes. The only differences come from building a model of increasing size.
On the other hand, the Symbolic/filt approach gains from pre-filtering. It drastically reduces the number of strategies to encode for the first two cases. For the last ones, there remains only one strategy to encode and check.

On the $1\ 1\ 1$ case, the Backward approach only needs to reach about half way---that is, to fix actions in states up to half of the depth of the model from the target states---to determine that there is no winning extension of the strategy that reaches the $\mathit{all\ defeated}$ states from the initial state.
For the other cases, it directly determines that there is no extension of the strategy that is winning in the initial state, thanks to its evaluation of the losing states. It does not need to extend the first considered strategy.

\paragraph{Comparison} The Partial approach does not handle the smallest model because it has to check the huge number of strategies to determine that there are no winning ones. On the other hand, the Partial/filt and Early/filt approaches only need pre-filtering to conclude.
The Early approach can also quickly determine that the formula is false because it just needs to check all possible actions in the initial state.
The symbolic approaches also perform well because the BDDs they compute remain very small.

The Backward approach is also quick because there is only one possible strategy for the two workers in states satisfying $\mathit{all\ defeated}$---doing nothing---,and there is no general strategy reaching these states from the initial one. The approach can thus directly conclude that the formula is false.

In conclusion, almost all approaches are comparable for the $1\ 2\ 2$ case and after because it is easy to show that the formula is false, except for the Partial approach that must check all possible strategies to reach this conclusion, and the Early one that has to check $16$ initial choices before concluding.

\section{Conclusion}\label{section:conclusion}
This paper described the \emph{backward} approach to solve the model-checking problem for $ATL_{ir}$, a logic reasoning about uniform memoryless strategies. The idea of the algorithm is to build the parts of winning uniform strategies from the target states. Unfortunately, the concept of building winning strategies from the ground up cannot be applied to greatest fixpoint computations, and the approach cannot be applied to $\CEW$ operators.
The approach has been implemented in a BDD-based framework with PyNuSMV, and has been experimentally compared to existing solutions. These experiments showed that the backward approach is competitive on the cases it can handle.

Pilecki et al.\ proposed a variant of their technique that only explores partial strategies defined along one single path of the system~\cite{Pilecki-Bednarczyk-others-14}. They showed that it is really efficient, but it is an \emph{incomplete} technique as it could miss some winning strategies. For this reason, the experiments did not take this variant into account. 

The limitation  to $\CEX$ and $\CEU$ operators can be overcome by mixing the approaches. They all compute the states satisfying strategic formulas. To evaluate the formula $\CEG{player}{\CEF{player}{win}}$ on the game repeated infinitely, we could compute the states satisfying $\CEF{player}{win}$ using the backward approach and evaluate the top-level strategic formula with another approach such as the partial one.

Other solutions have been proposed to solve problems similar to the model-checking problem of $ATL_{ir}$. Calta et al.\ proposed an algorithm to solve the problem of model checking $ATL_u$ formulas, a logic corresponding to $ATL_{ir}$ interpreted over sets of states of iCGS~\cite{Calta-Shkatov-others-10}. Nevertheless, their solution is not easily adapted to a BDD-based framework, so it is difficult to compare it with the backward approach.

Another solution has been proposed by Lomuscio and Raimondi~\cite{Lomuscio-Raimondi-06}. It works by enumerating all variants of the iCGS in which the agents act uniformly. Then the formula is satisfied if it is satisfied by one of these variants. The problem they solve is a bit different from $ATL_{ir}$ as, for the formula $\CEF{player}{win} \wedge \CEF{player}{lose}$ to be true, the same uniform strategy must be winning for both formulas, while $ATL_{ir}$ allows different strategies to witness the satisfaction. Also, their idea is very similar to the first solution of Busard et al.~\cite{Busard-Pecheur-others-13,Busard-Pecheur-others-15}, shown to be highly ineffective compared to the partial approach~\cite{Busard-Pecheur-others-14}.

Raskin et al.\ proposed an algorithm to check the existence of observation-based strategies for two-player turn-based games on graphs with $\omega$-regular objectives~\cite{Raskin-Chatterjee-others-07}. They are interested in the existence of winning observation-based strategies, that is, strategies with imperfect information and perfect recall. Their algorithm is based on antichains of state sets, that is, it works on the lattice of downward-closed subsets of states.
Another algorithm was recently proposed by Bozianu et al.~\cite{Bozianu-Dima-others-14}. It deals with the synthesis of a strategy with imperfect information and perfect recall for a single agent. As above, their algorithm works with antichains. But these solutions and the backward approach do not deal with the same logics: $ATL_{ir}$ is restricted to memoryless uniform strategies but can reason about several concurrent agents at the same time. On the other hand, the solutions above work with memory-full uniform strategies, but are limited to two-player turn-based games.

The verification of memoryless uniform strategies for reachability objectives is similar to strong planning under partial observability~\cite{Bertoli-Cimatti-others-06}. Strong planning under partial observability consists in finding a \emph{plan}---a sequence of actions to take, that may be conditioned by some observations on the system---that will surely reach a goal state. An additional constraint on these plans is that they will not reach the same \emph{belief state} twice---that is, they will not reach the same equivalence class twice.

The two settings are nevertheless different, and the link between memoryless uniform strategies and strong plans is not so clear. On one hand, we are interested in strategies for a set of agents that have different views of the system, while strong plans assume a unique observability relation. Furthermore, the restriction to memoryless strategies make them choose the same action for entire equivalence classes, while strong plans could encode some kind of finite-memory strategies---a plan could tell \emph{choose action $a$ then action $b$}, even if it ends up in the same belief state---but the additional constraint on the plans prevents them to do so. On the other hand, memoryless uniform strategies could reach the same equivalence class twice while still eventually reaching a target state.

Strong planning under partial observability has been extended to \emph{strong cyclic planning}, where the plans are defined as finite-state machines~\cite{Bertoli-Cimatti-others-06a}. These plans are successful if they reach a goal state, or if they loop again and again but still can reach a goal state after each loop. This setting is even further from uniform strategies than strong plans as they are not required to surely reach a goal state anymore.

\bibliographystyle{eptcs}
\bibliography{ref}

\begin{thebibliography}{10}
\providecommand{\bibitemdeclare}[2]{}
\providecommand{\surnamestart}{}
\providecommand{\surnameend}{}
\providecommand{\urlprefix}{Available at }
\providecommand{\url}[1]{\texttt{#1}}
\providecommand{\href}[2]{\texttt{#2}}
\providecommand{\urlalt}[2]{\href{#1}{#2}}
\providecommand{\doi}[1]{doi:\urlalt{http://dx.doi.org/#1}{#1}}
\providecommand{\bibinfo}[2]{#2}

\bibitemdeclare{article}{Alur-Henzinger-others-02}
\bibitem{Alur-Henzinger-others-02}
\bibinfo{author}{Rajeev \surnamestart Alur\surnameend},
  \bibinfo{author}{Thomas~A. \surnamestart Henzinger\surnameend} \&
  \bibinfo{author}{Orna \surnamestart Kupferman\surnameend}
  (\bibinfo{year}{2002}): \emph{\bibinfo{title}{Alternating-time temporal
  logic}}.
\newblock {\sl \bibinfo{journal}{J. ACM}}
  \bibinfo{volume}{49}(\bibinfo{number}{5}), pp. \bibinfo{pages}{672--713},
  \doi{10.1145/585265.585270}.

\bibitemdeclare{inproceedings}{Belardinelli-Lomuscio-others-17}
\bibitem{Belardinelli-Lomuscio-others-17}
\bibinfo{author}{Francesco \surnamestart Belardinelli\surnameend},
  \bibinfo{author}{Alessio \surnamestart Lomuscio\surnameend},
  \bibinfo{author}{Aniello \surnamestart Murano\surnameend} \&
  \bibinfo{author}{Sasha \surnamestart Rubin\surnameend}
  (\bibinfo{year}{2017}): \emph{\bibinfo{title}{Verification of Multi-agent
  Systems with Imperfect Information and Public Actions}}.
\newblock In: {\sl \bibinfo{booktitle}{Proceedings of AAMAS '17}}, pp.
  \bibinfo{pages}{1268--1276}.

\bibitemdeclare{inproceedings}{Berthon-Maubert-others-17}
\bibitem{Berthon-Maubert-others-17}
\bibinfo{author}{Rapha\"{e}l \surnamestart Berthon\surnameend},
  \bibinfo{author}{Bastien \surnamestart Maubert\surnameend} \&
  \bibinfo{author}{Aniello \surnamestart Murano\surnameend}
  (\bibinfo{year}{2017}): \emph{\bibinfo{title}{Decidability Results for ATL*
  with Imperfect Information and Perfect Recall}}.
\newblock In: {\sl \bibinfo{booktitle}{Proceedings of AAMAS '17}}, pp.
  \bibinfo{pages}{1250--1258}.

\bibitemdeclare{inproceedings}{Bertoli-Cimatti-others-06a}
\bibitem{Bertoli-Cimatti-others-06a}
\bibinfo{author}{Piergiorgio \surnamestart Bertoli\surnameend},
  \bibinfo{author}{Alessandro \surnamestart Cimatti\surnameend} \&
  \bibinfo{author}{Marco \surnamestart Pistore\surnameend}
  (\bibinfo{year}{2006}): \emph{\bibinfo{title}{Towards Strong Cyclic Planning
  Under Partial Observability}}.
\newblock In: {\sl \bibinfo{booktitle}{Proceedings of ICAPS'06}}, pp.
  \bibinfo{pages}{354--357}.

\bibitemdeclare{article}{Bertoli-Cimatti-others-06}
\bibitem{Bertoli-Cimatti-others-06}
\bibinfo{author}{Piergiorgio \surnamestart Bertoli\surnameend},
  \bibinfo{author}{Alessandro \surnamestart Cimatti\surnameend},
  \bibinfo{author}{Marco \surnamestart Roveri\surnameend} \&
  \bibinfo{author}{Paolo \surnamestart Traverso\surnameend}
  (\bibinfo{year}{2006}): \emph{\bibinfo{title}{Strong planning under partial
  observability}}.
\newblock {\sl \bibinfo{journal}{Artificial Intelligence}}
  \bibinfo{volume}{170}(\bibinfo{number}{4}), pp. \bibinfo{pages}{337 -- 384},
  \doi{10.1016/j.artint.2006.01.004}.

\bibitemdeclare{incollection}{Bozianu-Dima-others-14}
\bibitem{Bozianu-Dima-others-14}
\bibinfo{author}{Rodica \surnamestart Bozianu\surnameend},
  \bibinfo{author}{C{\u a}t{\u a}lin \surnamestart Dima\surnameend} \&
  \bibinfo{author}{Emmanuel \surnamestart Filiot\surnameend}
  (\bibinfo{year}{2014}): \emph{\bibinfo{title}{Safraless Synthesis for
  Epistemic Temporal Specifications}}.
\newblock In: {\sl \bibinfo{booktitle}{Computer Aided Verification}}, {\sl
  \bibinfo{series}{LNCS}} \bibinfo{volume}{8559}, \bibinfo{publisher}{Springer
  International Publishing}, pp. \bibinfo{pages}{441--456},
  \doi{10.1007/978-3-319-08867-9_29}.

\bibitemdeclare{article}{Bryant-86}
\bibitem{Bryant-86}
\bibinfo{author}{R.~E. \surnamestart Bryant\surnameend} (\bibinfo{year}{1986}):
  \emph{\bibinfo{title}{Graph-Based Algorithms for Boolean Function
  Manipulation}}.
\newblock {\sl \bibinfo{journal}{IEEE Transactions on Computers}}
  \bibinfo{volume}{C-35}(\bibinfo{number}{8}), pp. \bibinfo{pages}{677--691},
  \doi{10.1109/TC.1986.1676819}.

\bibitemdeclare{inproceedings}{Busard-Pecheur-13}
\bibitem{Busard-Pecheur-13}
\bibinfo{author}{Simon \surnamestart Busard\surnameend} \&
  \bibinfo{author}{Charles \surnamestart Pecheur\surnameend}
  (\bibinfo{year}{2013}): \emph{\bibinfo{title}{PyNuSMV: NuSMV as a Python
  Library}}.
\newblock In: {\sl \bibinfo{booktitle}{Proceedings of NFM 2013}}, pp.
  \bibinfo{pages}{453--458}, \doi{10.1007/978-3-642-38088-4_33}.

\bibitemdeclare{inproceedings}{Busard-Pecheur-others-13}
\bibitem{Busard-Pecheur-others-13}
\bibinfo{author}{Simon \surnamestart Busard\surnameend},
  \bibinfo{author}{Charles \surnamestart Pecheur\surnameend},
  \bibinfo{author}{Hongyang \surnamestart Qu\surnameend} \&
  \bibinfo{author}{Franco \surnamestart Raimondi\surnameend}
  (\bibinfo{year}{2013}): \emph{\bibinfo{title}{Reasoning about Strategies
  under Partial Observability and Fairness Constraints}}.
\newblock In: {\sl \bibinfo{booktitle}{Proceedings of SR 2013}}, pp.
  \bibinfo{pages}{71--79}, \doi{10.4204/EPTCS.112.12}.

\bibitemdeclare{incollection}{Busard-Pecheur-others-14}
\bibitem{Busard-Pecheur-others-14}
\bibinfo{author}{Simon \surnamestart Busard\surnameend},
  \bibinfo{author}{Charles \surnamestart Pecheur\surnameend},
  \bibinfo{author}{Hongyang \surnamestart Qu\surnameend} \&
  \bibinfo{author}{Franco \surnamestart Raimondi\surnameend}
  (\bibinfo{year}{2014}): \emph{\bibinfo{title}{Improving the Model Checking of
  Strategies under Partial Observability and Fairness Constraints}}.
\newblock In: {\sl \bibinfo{booktitle}{Formal Methods and Software
  Engineering}}, {\sl \bibinfo{series}{LNCS}} \bibinfo{volume}{8829},
  \bibinfo{publisher}{Springer International Publishing}, pp.
  \bibinfo{pages}{27--42}, \doi{10.1007/978-3-319-11737-9_3}.

\bibitemdeclare{article}{Busard-Pecheur-others-15}
\bibitem{Busard-Pecheur-others-15}
\bibinfo{author}{Simon \surnamestart Busard\surnameend},
  \bibinfo{author}{Charles \surnamestart Pecheur\surnameend},
  \bibinfo{author}{Hongyang \surnamestart Qu\surnameend} \&
  \bibinfo{author}{Franco \surnamestart Raimondi\surnameend}
  (\bibinfo{year}{2015}): \emph{\bibinfo{title}{Reasoning about memoryless
  strategies under partial observability and unconditional fairness
  constraints}}.
\newblock {\sl \bibinfo{journal}{Information and Computation}}
  \bibinfo{volume}{242}, pp. \bibinfo{pages}{128 -- 156},
  \doi{10.1016/j.ic.2015.03.014}.

\bibitemdeclare{incollection}{Calta-Shkatov-others-10}
\bibitem{Calta-Shkatov-others-10}
\bibinfo{author}{Jan \surnamestart Calta\surnameend}, \bibinfo{author}{Dmitry
  \surnamestart Shkatov\surnameend} \& \bibinfo{author}{Holger \surnamestart
  Schlingloff\surnameend} (\bibinfo{year}{2010}): \emph{\bibinfo{title}{Finding
  Uniform Strategies for Multi-agent Systems}}.
\newblock In: {\sl \bibinfo{booktitle}{Computational Logic in Multi-Agent
  Systems}}, {\sl \bibinfo{series}{LNCS}} \bibinfo{volume}{6245},
  \bibinfo{publisher}{Springer}, pp. \bibinfo{pages}{135--152},
  \doi{10.1007/978-3-642-14977-1\_12}.

\bibitemdeclare{incollection}{Cimatti-Clarke-others-02}
\bibitem{Cimatti-Clarke-others-02}
\bibinfo{author}{Alessandro \surnamestart Cimatti\surnameend},
  \bibinfo{author}{Edmund \surnamestart Clarke\surnameend},
  \bibinfo{author}{Enrico \surnamestart Giunchiglia\surnameend},
  \bibinfo{author}{Fausto \surnamestart Giunchiglia\surnameend},
  \bibinfo{author}{Marco \surnamestart Pistore\surnameend},
  \bibinfo{author}{Marco \surnamestart Roveri\surnameend},
  \bibinfo{author}{Roberto \surnamestart Sebastiani\surnameend} \&
  \bibinfo{author}{Armando \surnamestart Tacchella\surnameend}
  (\bibinfo{year}{2002}): \emph{\bibinfo{title}{{NuSMV} 2: An OpenSource Tool
  for Symbolic Model Checking}}.
\newblock In: {\sl \bibinfo{booktitle}{Computer Aided Verification}},
  \bibinfo{publisher}{Springer}, pp. \bibinfo{pages}{359--364},
  \doi{10.1007/3-540-45657-0_29}.

\bibitemdeclare{inproceedings}{Dastani-Jamroga-10}
\bibitem{Dastani-Jamroga-10}
\bibinfo{author}{Mehdi \surnamestart Dastani\surnameend} \&
  \bibinfo{author}{Wojciech \surnamestart Jamroga\surnameend}
  (\bibinfo{year}{2010}): \emph{\bibinfo{title}{Reasoning about strategies of
  multi-agent programs}}.
\newblock In: {\sl \bibinfo{booktitle}{Proceedings of AAMAS 10}}, pp.
  \bibinfo{pages}{997--1004}.

\bibitemdeclare{article}{Dima-Tiplea-11}
\bibitem{Dima-Tiplea-11}
\bibinfo{author}{Catalin \surnamestart Dima\surnameend} \&
  \bibinfo{author}{Ferucio~Laurentiu \surnamestart Tiplea\surnameend}
  (\bibinfo{year}{2011}): \emph{\bibinfo{title}{Model-checking {ATL} under
  Imperfect Information and Perfect Recall Semantics is Undecidable}}.
\newblock {\sl \bibinfo{journal}{CoRR}} \bibinfo{volume}{abs/1102.4225}.
\newblock \urlprefix\url{http://arxiv.org/abs/1102.4225}.

\bibitemdeclare{inproceedings}{Huang-Meyden-14}
\bibitem{Huang-Meyden-14}
\bibinfo{author}{Xiaowei \surnamestart Huang\surnameend} \&
  \bibinfo{author}{Ron \surnamestart van~der Meyden\surnameend}
  (\bibinfo{year}{2014}): \emph{\bibinfo{title}{Symbolic Model Checking
  Epistemic Strategy Logic}}.
\newblock In: {\sl \bibinfo{booktitle}{Proceedings of the Twenty-Eighth {AAAI}
  Conference on Artificial Intelligence}}, pp. \bibinfo{pages}{1426--1432}.

\bibitemdeclare{inproceedings}{Jamroga-Dix-06}
\bibitem{Jamroga-Dix-06}
\bibinfo{author}{Wojciech \surnamestart Jamroga\surnameend} \&
  \bibinfo{author}{J\"urgen \surnamestart Dix\surnameend}
  (\bibinfo{year}{2006}): \emph{\bibinfo{title}{Model Checking Abilities under
  Incomplete Information Is Indeed {$\Delta^P_2$}-complete}}.
\newblock In: {\sl \bibinfo{booktitle}{EUMAS'06}}.

\bibitemdeclare{article}{Jamroga-Hoek-04}
\bibitem{Jamroga-Hoek-04}
\bibinfo{author}{Wojciech \surnamestart Jamroga\surnameend} \&
  \bibinfo{author}{Wiebe \surnamestart van~der Hoek\surnameend}
  (\bibinfo{year}{2004}): \emph{\bibinfo{title}{Agents that Know How to Play}}.
\newblock {\sl \bibinfo{journal}{Fundamenta Informaticae}}
  \bibinfo{volume}{Volume 63}(\bibinfo{number}{2}), pp.
  \bibinfo{pages}{185--219}.

\bibitemdeclare{inproceedings}{Lomuscio-Raimondi-06}
\bibitem{Lomuscio-Raimondi-06}
\bibinfo{author}{Alessio \surnamestart Lomuscio\surnameend} \&
  \bibinfo{author}{Franco \surnamestart Raimondi\surnameend}
  (\bibinfo{year}{2006}): \emph{\bibinfo{title}{Model checking knowledge,
  strategies, and games in multi-agent systems}}.
\newblock In: {\sl \bibinfo{booktitle}{{AAMAS} 2006, Hakodate, Japan, May 8-12,
  2006}}, pp. \bibinfo{pages}{161--168}, \doi{10.1145/1160633.1160660}.

\bibitemdeclare{article}{Peterson-Reif-others-02}
\bibitem{Peterson-Reif-others-02}
\bibinfo{author}{G.~\surnamestart Peterson\surnameend},
  \bibinfo{author}{J.~\surnamestart Reif\surnameend} \&
  \bibinfo{author}{S.~\surnamestart Azhar\surnameend} (\bibinfo{year}{2002}):
  \emph{\bibinfo{title}{Decision algorithms for multiplayer noncooperative
  games of incomplete information}}.
\newblock {\sl \bibinfo{journal}{Computers and Mathematics with Applications}}
  \bibinfo{volume}{43}(\bibinfo{number}{1}), pp. \bibinfo{pages}{179 -- 206},
  \doi{10.1016/S0898-1221(01)00282-6}.

\bibitemdeclare{incollection}{Pilecki-Bednarczyk-others-14}
\bibitem{Pilecki-Bednarczyk-others-14}
\bibinfo{author}{Jerzy \surnamestart Pilecki\surnameend},
  \bibinfo{author}{Marek~A. \surnamestart Bednarczyk\surnameend} \&
  \bibinfo{author}{Wojciech \surnamestart Jamroga\surnameend}
  (\bibinfo{year}{2014}): \emph{\bibinfo{title}{Synthesis and Verification of
  Uniform Strategies for Multi-agent Systems}}.
\newblock In: {\sl \bibinfo{booktitle}{Computational Logic in Multi-Agent
  Systems}}, {\sl \bibinfo{series}{LNCS}} \bibinfo{volume}{8624},
  \bibinfo{publisher}{Springer International Publishing}, pp.
  \bibinfo{pages}{166--182}, \doi{10.1007/978-3-319-09764-0_11}.

\bibitemdeclare{inbook}{Ramanujam-Simon-10}
\bibitem{Ramanujam-Simon-10}
\bibinfo{author}{R.~\surnamestart Ramanujam\surnameend} \&
  \bibinfo{author}{Sunil \surnamestart Simon\surnameend}
  (\bibinfo{year}{2010}): \emph{\bibinfo{title}{A Communication Based Model for
  Games of Imperfect Information}}, pp. \bibinfo{pages}{509--523}.
\newblock \bibinfo{publisher}{Springer Berlin Heidelberg},
  \bibinfo{address}{Berlin, Heidelberg}, \doi{10.1007/978-3-642-15375-4_35}.

\bibitemdeclare{article}{Raskin-Chatterjee-others-07}
\bibitem{Raskin-Chatterjee-others-07}
\bibinfo{author}{Jean{-}Fran{\c{c}}ois \surnamestart Raskin\surnameend},
  \bibinfo{author}{Krishnendu \surnamestart Chatterjee\surnameend},
  \bibinfo{author}{Laurent \surnamestart Doyen\surnameend} \&
  \bibinfo{author}{Thomas~A. \surnamestart Henzinger\surnameend}
  (\bibinfo{year}{2007}): \emph{\bibinfo{title}{Algorithms for Omega-Regular
  Games with Imperfect Information}}.
\newblock {\sl \bibinfo{journal}{Logical Methods in Computer Science}}
  \bibinfo{volume}{3}(\bibinfo{number}{3}), \doi{10.2168/LMCS-3(3:4)2007}.

\bibitemdeclare{article}{Schobbens-04}
\bibitem{Schobbens-04}
\bibinfo{author}{Pierre-Yves \surnamestart Schobbens\surnameend}
  (\bibinfo{year}{2004}): \emph{\bibinfo{title}{Alternating-time logic with
  imperfect recall}}.
\newblock {\sl \bibinfo{journal}{Electronic Notes in Theoretical Computer
  Science}} \bibinfo{volume}{85}(\bibinfo{number}{2}), pp. \bibinfo{pages}{82
  -- 93}, \doi{10.1016/S1571-0661(05)82604-0}.

\end{thebibliography}
\end{document}